\documentclass[%
 reprint,
 amsmath,amssymb,
 aps,
]{revtex4-2}
\usepackage{siunitx}
\usepackage{soul}
\usepackage{times}
\usepackage{booktabs}
\usepackage{threeparttable}
\usepackage{graphicx}
\usepackage{dcolumn}
\usepackage{bm}

\begin{document}

\preprint{APS/123-QED}


\title{Stabilizing by steering: Enhancing bacterial motility by non-uniform diffusiophoresis}

\author{Viet Sang Doan}
\thanks{These authors contributed equally to this work.}
\affiliation{Department of Mechanical and Aerospace Engineering, University at Buffalo, The State University of New York, Buffalo, New York 14260, USA}
\author{Ali Nikkhah}
\thanks{These authors contributed equally to this work.}
\affiliation{Department of Mechanical and Aerospace Engineering, University at Buffalo, The State University of New York, Buffalo, New York 14260, USA}
\author{Sangwoo shin}
\affiliation{Department of Mechanical and Aerospace Engineering, University at Buffalo, The State University of New York, Buffalo, New York 14260, USA}

\date{\today}
\begin{abstract}
Bacteria often traverse confined spaces to perform critical functions in symbiosis, infection, drug delivery, and soil bioremediation. While the canonical run-and-tumble strategy enables exploration, its reliance on stochastic reorientation limits efficiency under confinement. We show that salt gradients can physically steer \textit{Pseudomonas putida} by biasing their runs toward salt through asymmetric diffusiophoretic forces. These gradients impose a torque strong enough to overcome Brownian rotation, aligning cells along the gradient and producing straighter, more persistent motion. This response persists in a $\Delta$\textit{cheA} mutant, demonstrating that canonical CheA-dependent chemotaxis is not required. We further show that when toxic organic contaminants are present, salt gradients promote bacterial dispersion toward the source despite reduced motility, suggesting potential applications in bacterial delivery for bioremediation. This work uncovers a previously unrecognized mechanism by which salt gradients direct bacterial motility, revealing a physical route to control microbial transport in complex environments.
\end{abstract}

\maketitle



Bioaugmentation is a soil bioremediation strategy in which cultured bacteria are introduced into contaminated soil to degrade pollutants \citep{vogel1996bioaugmentation}. Among numerous factors that determine the process efficacy, 
a critical prerequisite that is often challenging to achieve is to disperse the cells to the pollutant source, particularly to low-permeability zones that are abundant in heterogeneous, unsaturated soil \citep{lyon2012bioaugmentation}. Successful bioaugmentation thus requires directing decomposer bacteria to the target soil micropores deep in the subsurface where the contaminants persist \citep{pieper2000engineering,yang2018mechanisms}. Bacteria can be carried through permeable regions by hydrodynamic dispersion, but transport into less permeable regions depends primarily on active motility or Brownian motion. Because these regions can retain substantial amounts of contaminants, limited cell access can limit the effectiveness of remediation \citep{sturman1995engineering,alexander1999biodegradation}.

The fact that decomposer bacteria can thrive on toxic contaminants makes them naturally chemotactic toward these contaminants \citep{harwood1984aromatic,law2003bacterial,giri2021progress}. In fact, it is well known that chemotaxis directs cells to pollutant-rich regions and increases bacterial residence in porous media containing residual contaminants \citep{duffy1997residence,adadevoh2016chemotaxis,adadevoh2018chemotaxis,de2021chemotaxis}. Nevertheless, the inherently stochastic nature of flagellar motility (\textit{e.g.}, run-and-tumble) may limit navigation efficiency \citep{berg1990chemotaxis,ford2007role}; thus, rather deterministic transport mechanisms may offer more reliable targeting. Here, we hypothesize that \textit{diffusiophoresis} could provide this deterministic guidance, enabling more directed bacterial migration toward contaminant-rich regions.

Diffusiophoresis describes the directed motion of colloidal particles along solute gradients, where the particle velocity can be precisely controlled by manipulating solute gradients \citep{ault2024physicochemical}. 
A number of recent studies have shown that diffusiophoresis can be used to control colloidal transport in confined spaces, \textit{e.g.}, porous media \citep{park2021microfluidic,doan2021confinement,somasundar2023diffusiophoretic,sambamoorthy2023diffusiophoresis,jotkar2024diffusiophoresis,jotkar2024impact,alipour2026diffusiophoretic} and impervious dead-ends \citep{kar2015enhanced,shin2016size,ault2017diffusiophoresis,shin2018cleaning,tan2021two,shi2021droplet,duong2026salt}, as such regions typically display diffusion-limited environments. 
Also, the native surface charge of cell membranes makes bacteria susceptible to experiencing diffusiophoresis \citep{lee2018diffusiophoretic,doan2020trace,shim2021co}, although it remains elusive how diffusiophoresis affects flagellar motility.
Motivated by these prior studies, here we investigate the impact of salt-gradient-driven diffusiophoresis on the motility of soil bacterium \textit{Pseudomonas putida} F1 and demonstrate the use of diffusiophoresis in enhancing chemotactic dispersion toward non-aqueous phase liquid (NAPL) in confined geometries. 

\section*{Results}
\subsection*{Salt Gradients Induce Directional Cell Migration}
To examine the motility of \textit{P. putida} F1 under the influence of salt gradients, we used a transparent microfluidic device consisting of two parallel flow channels connected by a narrow pore (Figure 1a). One end of the pore is sealed with a thin hydrogel membrane (polyethylene glycol diacrylate) to suppress undesired convective flows \citep{paustian2013microfluidic}. The membrane allows the diffusion of solute molecules, enabling the formation of stable salt gradients. Additional details on the fabrication process are provided in the Supplemental Material (SM) and also in our previous work \citep{doan2024diffusiophoresis}.

\begin{figure}[t!]
	\centering
	\includegraphics[width=1.00\linewidth]{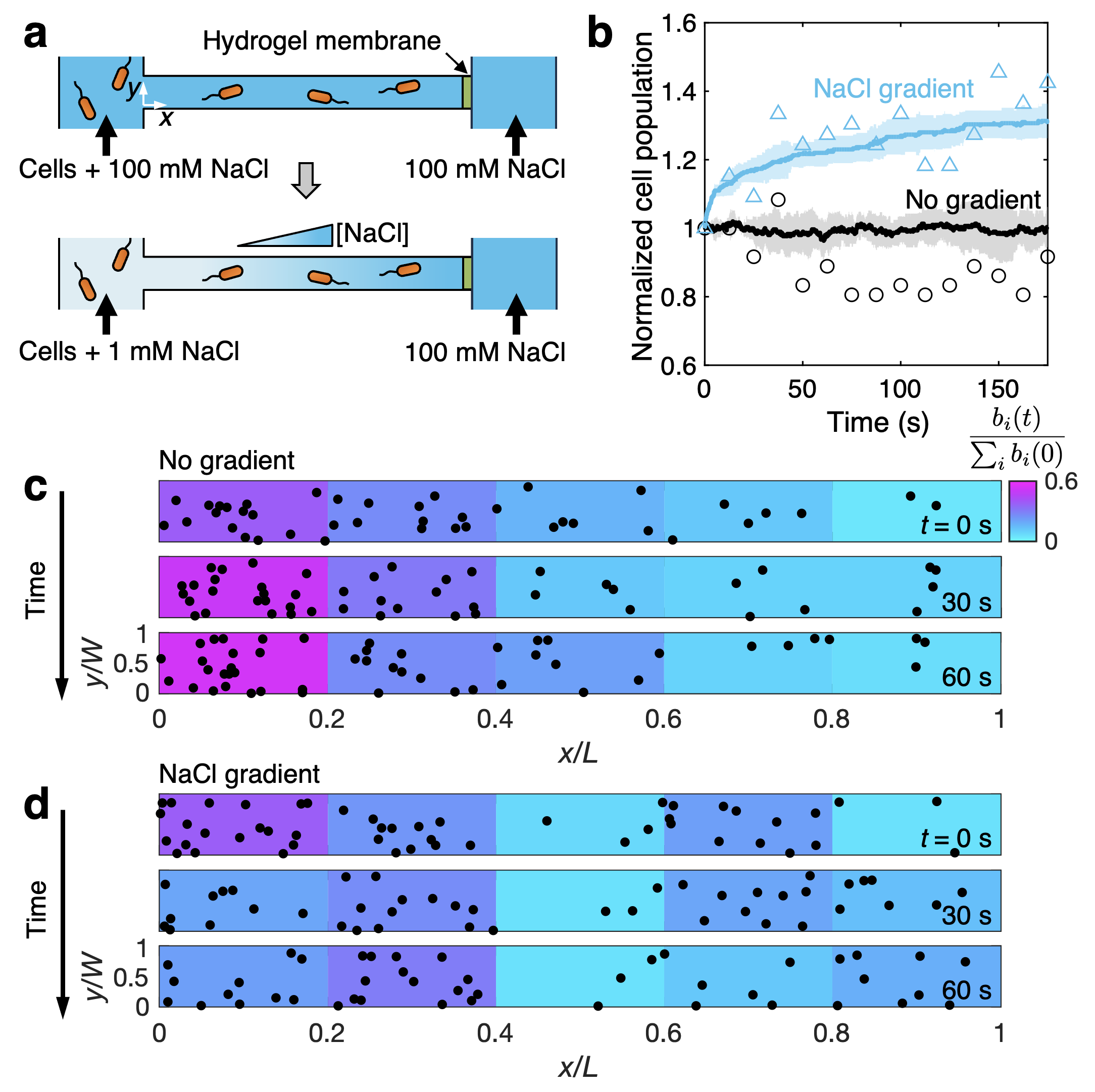}
	\caption{
		\textbf{\textit{P. putida} are attracted to higher salinity.}	
		(a) A microfluidic assay to evaluate cells under NaCl gradients. 
		(b) Change in the total cell population normalized by the initial cell population over time in the presence (blue) and absence (black) of NaCl gradients. Symbols represent experimental data (blue triangles for NaCl gradients, black circles for no gradient) while solid lines represent numerical simulations. Shaded regions denote the standard deviation across three simulations.
		(c,d) Local distribution of cells over time (c) without (Movie S1) or (d) with NaCl gradients (Movie S2). The color code represents the number of cells in the $i$-th bin, $b_i(t)$, normalized by the total number of cells from the beginning, $\sum_i b_i(0)$. Although the heat map only shows one experiment here, the observed patterns were reproduced across multiple replicates, with similar trends
	}
\end{figure}

Initially, cells suspended in a 10$\times$-diluted random motility buffer (11.2~g/L K$_2$HPO$_4$, 4.8~g/L KH$_2$PO$_4$, and 0.029~g/L EDTA) supplemented with 0.1~M NaCl are introduced into the open-sided flow channel (left channel in Figure 1a), while the same buffer containing 0.1~M NaCl is simultaneously injected into the gel-sided flow channel (right channel). Bacteria from the left flow channel swim into the pore, eventually achieving an even distribution across the pore after one hour (upper panel in Figure~1a). Then, a cell suspension at reduced salt concentration (1~mM NaCl with dilute motility buffer) is injected into the left flow channel to create salt gradients across the pore (lower panel in Figure 1a).  

Upon the introduction of NaCl gradients, we observe that the overall cell population in the pore changes over time. As presented in Figure 1b, the total number of cells in the pore gradually increases in the presence of NaCl gradients, whereas the cell population remains nearly unchanged when the NaCl concentration is uniform throughout the channel. Moreover, not only does the total cell population increase, but also their spatial distribution changes, as shown in Figures 1c,d; the color code represents the local number of cells normalized by the total cell number at $t=0$. While the overall cell distribution remains more or less the same in the absence of NaCl gradients (Figure 1c, Movie S1), with NaCl gradients the cells tend to locate deeper into the channel toward higher salt concentration (Figure 1d, Movie S2). These results suggest that salinity gradients can attract the bacteria into the deep pores and promote their dispersion toward higher salinity.

\begin{figure}[t!]
	\centering
	\includegraphics[width=1.00\linewidth]{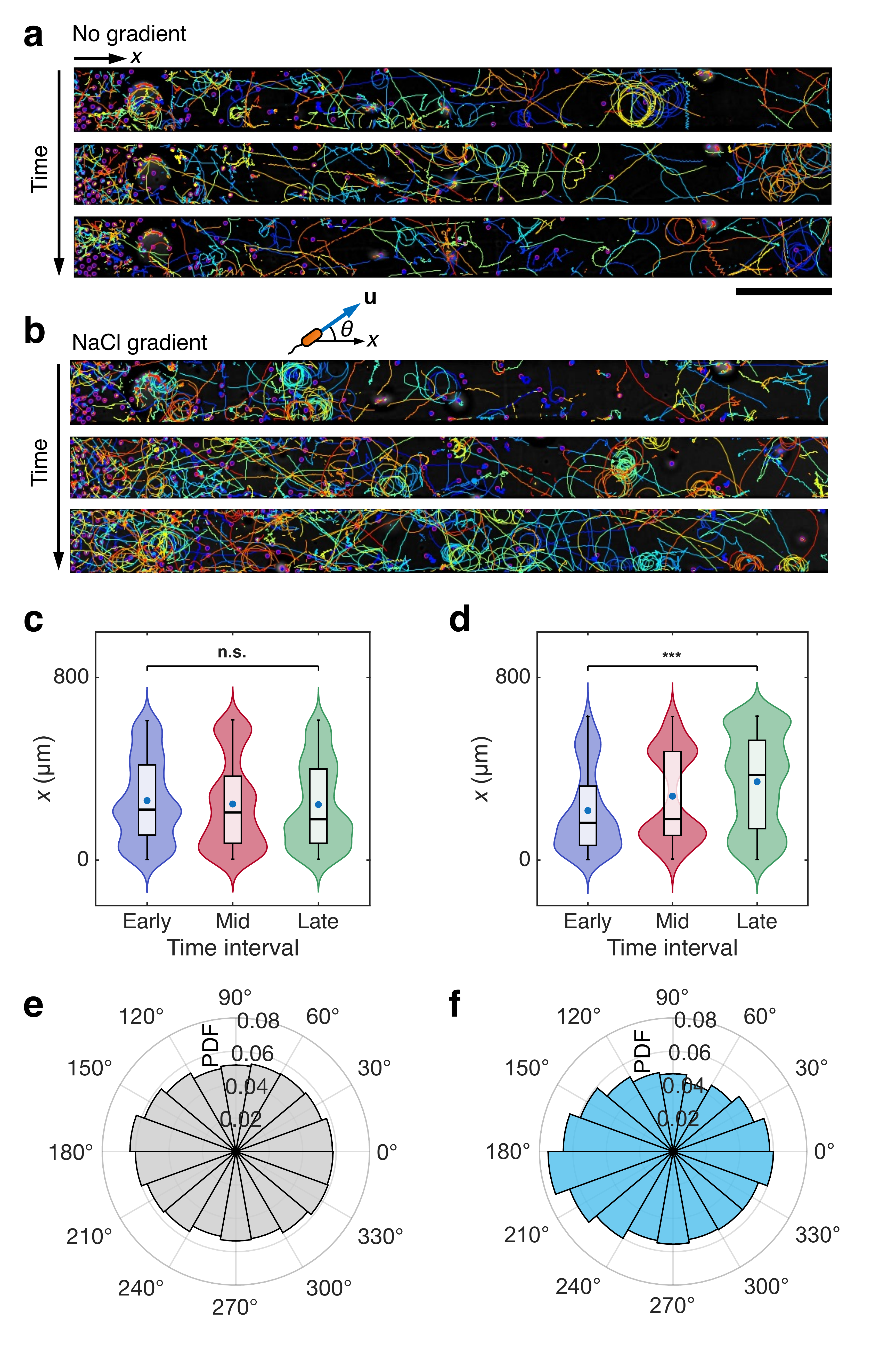}
	\caption{
		\textbf{\textit{P. putida} aligns along salt gradients.}	
		(a,b) Unfiltered cell trajectories recorded over 25 s of duration for different times (0--25, 75--100, 150--175~s) in the (a) absence and (b) presence of NaCl gradients.  (c,d) Distribution of cell mean positions along the $x$ axis at successive 25 s time intervals (Early: 0--25~s, Mid: 75--100~s, Late: 150--175~s) in the (c) absence and (d) presence of NaCl gradients. The white box in (c,d) indicates interquartile range with median shown as a stripe and mean as a dot. (e,f) Probability density function distribution of cell's directions of instantaneous velocity vectors in the (e) absence and (f) presence of NaCl gradients. Scale bar in (a) is 100~$\mu$m. 
	} 
	\label{fig:division}
\end{figure}

\subsection*{Directional alignment and biased motility under salt gradients}
To better understand the observed biased behavior, we evaluated the motility of individual cells by analyzing their trajectories. Unfiltered sample trajectories with and without NaCl gradients at different time windows over a period of 25~s are presented in Figures 2a,b, which also display more cell trajectories found deep in the pore over time under NaCl gradients (Figure~2b). 

\begin{figure*}[t!]
	\makebox[\linewidth][c]{\includegraphics[width=1.0\linewidth]{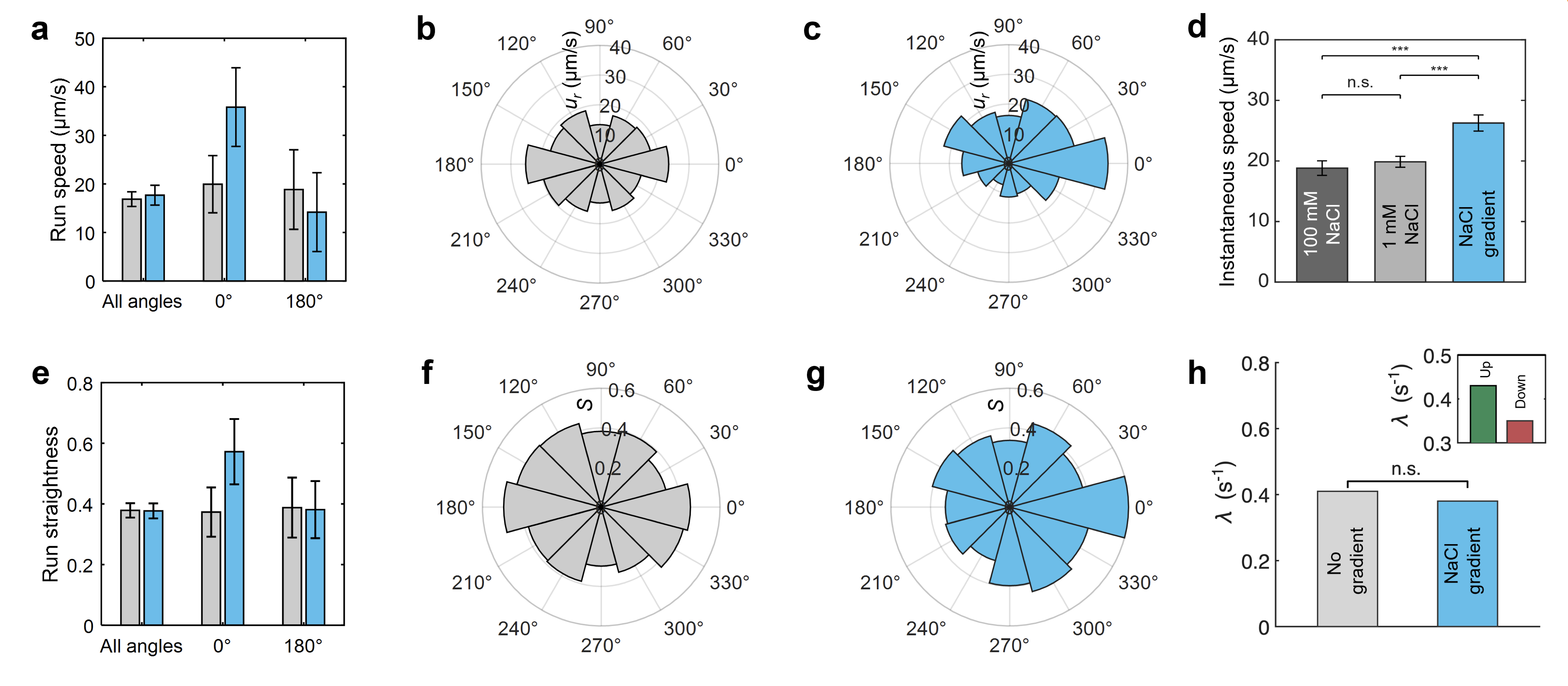}}    \centering
	\caption{
		\textbf{\textit{P. putida} run faster and straighter toward salt.}	
		(a) Comparison of run speeds averaged over the entire angles, $0^\circ~(-15^\circ<\theta<15^\circ)$, and $180^\circ~(165^\circ<\theta<195^\circ)$. 
		(b,c) Distributions of average run speed in the (b) absence and (c) presence of NaCl gradients. (d) Instantaneous run speed in the absence and presence of NaCl gradients 
		(e) Comparison of average run straightness for entire angles, $0^\circ~(-15^\circ<\theta<15^\circ)$, and $180^\circ~(165^\circ<\theta<195^\circ)$. 
		(f,g) Distributions of average run straightness in the (f) absence and (g) presence of NaCl gradients. (h) Tumble rates measured in the absence and presence of NaCl gradients. The inset compares tumble rates for runs oriented up gradient and down gradient.
		Error bars in (a, d, e) represent standard deviation.
	}
\end{figure*}

To quantitatively determine whether more cells move to the end of the pore, we analyzed the distribution of cell positions over time (Figure~2c,d). In the absence of NaCl gradients, the distribution remains spatially consistent, with no progressive shift in the position and distribution (Figure~2c). We did not observe any temporal change between early (0--25~s) and late (150--175~s) segments ($\Delta x = -17.43 \pm 18.9$~\textmu m, $p = 0.36$, $n = 375$ trajectories) \citep{sullivan2012using}. In contrast, under NaCl gradients, the distribution shifts progressively deeper into the pore, as reflected by increases in the median and a broader distribution at later times extending toward larger $x$ values (deeper into the pore), as shown in Figure 2d. The median position is increased from $216.9 \pm 15.1$~\textmu m to $343.2 \pm 19.6$~\textmu m ($\Delta x = 126.30 \pm 24.7$~\textmu m, $p$ $<$ 0.0001, $n = 255$ trajectories). More statistical details can be found in Table S1 (SM).

Upon analyzing the trajectories, we immediately identify that the cells tend to be more aligned along the salt gradient, where Figures 2e,f show the probability distribution of angle $\theta$ between the cell's instantaneous velocity vector $\bm{u}$ and the pore axis ($+x$ direction). The slightly reduced distribution in the lateral direction ($\theta \rightarrow \pm90^\circ$) in the absence of NaCl gradients (Figure 2e) is likely due to the lateral confinement imposed by the slender geometry of the pore (width $\approx$~65~\textmu m).
In the presence of NaCl gradients, the noticeable alignment toward the pore entrance ($\theta \rightarrow 180^\circ$), i.e., away from the salt, is due to the run-and-reverse motility of \textit{P. putida}, which is one of the pronounced motilities of lophotrichous \textit{P. putida} \citep{harwood1989flagellation,alirezaeizanjani2020chemotaxis,thormann2022wrapped}. Regardless of whether the cells are headed toward or away from the salt, the cells are aligned more along the pore direction (Figure~2f).

Next, we analyzed their mean run speed, i.e., the average speed of the cell between successive tumble events.
Here, we define `tumble' as any sudden change in the cell speed and direction between two successive runs (Figure~S1, SM).
After processing over 1,400 runs combined, we did not observe any noticeable difference in the average run speed between the control group (uniform concentration) and the NaCl gradient cases (Figure 3a). However, when we plot the run speed in their corresponding directions (Figures 3b,c), we notice that there is a significant improvement in the run speed toward the salt ($\theta \rightarrow 0^\circ$), and at the same time, the run speed decreases in the direction away from the salt ($\theta \rightarrow 180^\circ$). 

We note that the tumble rates did not differ significantly between the control and NaCl gradient cases (Figure~3h; Figure~S2, SM), suggesting that NaCl gradients do not interfere with chemoreceptors \citep{qi1989salt}.
Further testing with \mbox{$\Delta$\textit{cheA}} mutant, which lacks the ability to relay chemoreceptor input to the flagellar motor and thus are unable to bias their swimming in response to chemical gradients \citep{stock1988chea}, also yields similar conclusion. \mbox{$\Delta$\textit{cheA}} cells still migrate deeper into the pore, as reflected by increases in the median and a broader cell distribution at later times extending deeper into the pore (Figure~S3a,b, SM), and also exhibits the same directional bias in run speed as the wild type (Figure~S3c,d, SM), indicating that the observed response is not driven by chemotaxis.

While the observed bias in the run speed may be speculated as due to the additional diffusiophoretic drift toward salt, the speed change in fact cannot be explained by diffusiophoresis alone. Under current conditions, the diffusiophoretic velocity, $u_d \approx \mathcal{M}_d/\ell$ where $\mathcal{M}_d=\mathcal{O}(100~\mathrm{\mu m^2/s})$ is the diffusiophoretic mobility \citep{doan2020trace} and $\ell=\mathcal{O}(100~\mathrm{\mu m})$ is the characteristic length scale of the salt gradient, is estimated to be no more than 1~$\mu$m/s, which is an order of magnitude smaller than the measured speed change.
Another possibility for the observed bias may be due to chemokinesis, which is a non-directional response in which cells modulate their swimming speed in response to local chemical concentration \citep{gao2021coral}. This behavior has been reported in uniform attractant concentrations without gradients, e.g., in \textit{Rhodobacter sphaeroides} \textit{Escherichia coli} and \textit{Azospirillum brasilense} \citep{packer1994chemokinetic,deepika2015variation,zhulin1993motility}.
We analyzed the instantaneous run speed of the bacteria and found no significant differences between 1~mM and 100~mM NaCl concentrations. Nonetheless, the difference was markedly significant ($p$~$<$~0.001) under the salt gradients, as illustrated in Figure 3d. More statistical analysis can be found in Table~S2 (SM). Therefore, we conclude that the observed bias in the run speed cannot be fully explained by chemokinesis. 

Interestingly, as we observe how cells run by analyzing the straightness of each run, we notice that cells tend to run straighter toward the salt (Figures 3e-g). 
The run straightness, $\mathcal{S} = | \bm{x}_k - \bm{x}_1|/\sum_{j=1}^{k-1} (|\bm{x}_{j+1}|-|\bm{x}_{j}|)$, where $\bm{x}_j$ is the cell position at $j$-th frame within a single run, which is defined as the ratio of the end-to-end distance to the total distance the cell has traveled within a run, represents an effective tortuosity of runs \citep{benhamou2004reliably,kamdar2022colloidal}. 
As shown in Figure 3g, the run straightness is noticeably improved when swimming toward the salt. Similar directional bias in the run straightness was also observed in \mbox{$\Delta$\textit{cheA}} mutant. This behavior was unexpected since run straightness is mainly affected by Brownian rotation or the inherent asymmetry of the cell, both of which are sensitive to the cell morphology \citep{hyon2012wiggling}.

\textit{P. putida} is lophotrichous, where multiple flagella (length $f \sim 5~\mu$m, thickness $\sim 20$~nm) emerge from one pole of the cell body (body length $a \sim 1.5~ \mu$m, body width $b \sim 0.8~\mu$m) \citep{harwood1989flagellation,tokarova2021patterns}.
Therefore, when \textit{P. putida} propels forward the flagella tuft bundles up and forms a long, slender helix that is attached to the rod-like body where the body and the flagellar bundle are spatially segregated \citep{park2024bundling}. Given the asymmetry of this geometry, we hypothesize that the improved straightness toward the salt is due to \textit{non-uniform diffusiophoresis} acting distinctly on the cell body and flagellar bundle.

\subsection*{Non-uniform diffusiophoresis steers cells}
Under the influence of salt gradients, the surface charge of the cell membrane triggers bacterial diffusiophoresis regardless of their motility \citep{doan2020trace}. However, because there is a significant difference in the size and surface charge between the body (mainly consists of lipopolysaccharides and phospholipids) and the flagella (consist of flagellin), we expect that diffusiophoresis will not be occurring uniformly over the entire cell owing to the size- and charge-dependent nature of diffusiophoresis \citep{PrieveJFM1984,prieve1987diffusiophoresis,shin2016size,doan2023shape,lee2023role,akdeniz2023diffusiophoresis}. As the cell body is larger and more charged than flagella, the body is expected to experience stronger diffusiophoresis. We confirm that the surface charge of the flagella is much smaller than the cell body, where the measured zeta potential of flagella isolated from the cell body is $-0.6\pm 1.0$~mV whereas the entire body is $-16.3\pm 0.6$~mV (SM).
Considering the size and surface charge, the diffusiophoretic mobility of the body and flagellar bundle are estimated to be $\mathcal{M}_d^\textrm{body}=99.2~\mu\mathrm{m^2/s}$ and $\mathcal{M}_d^\textrm{flag}=2.8~\mu\mathrm{m^2/s}$, respectively \citep{PrieveJFM1984,shin2016size}.

\begin{figure*}[t!]
	\centering
	\includegraphics[width=1.00\linewidth]{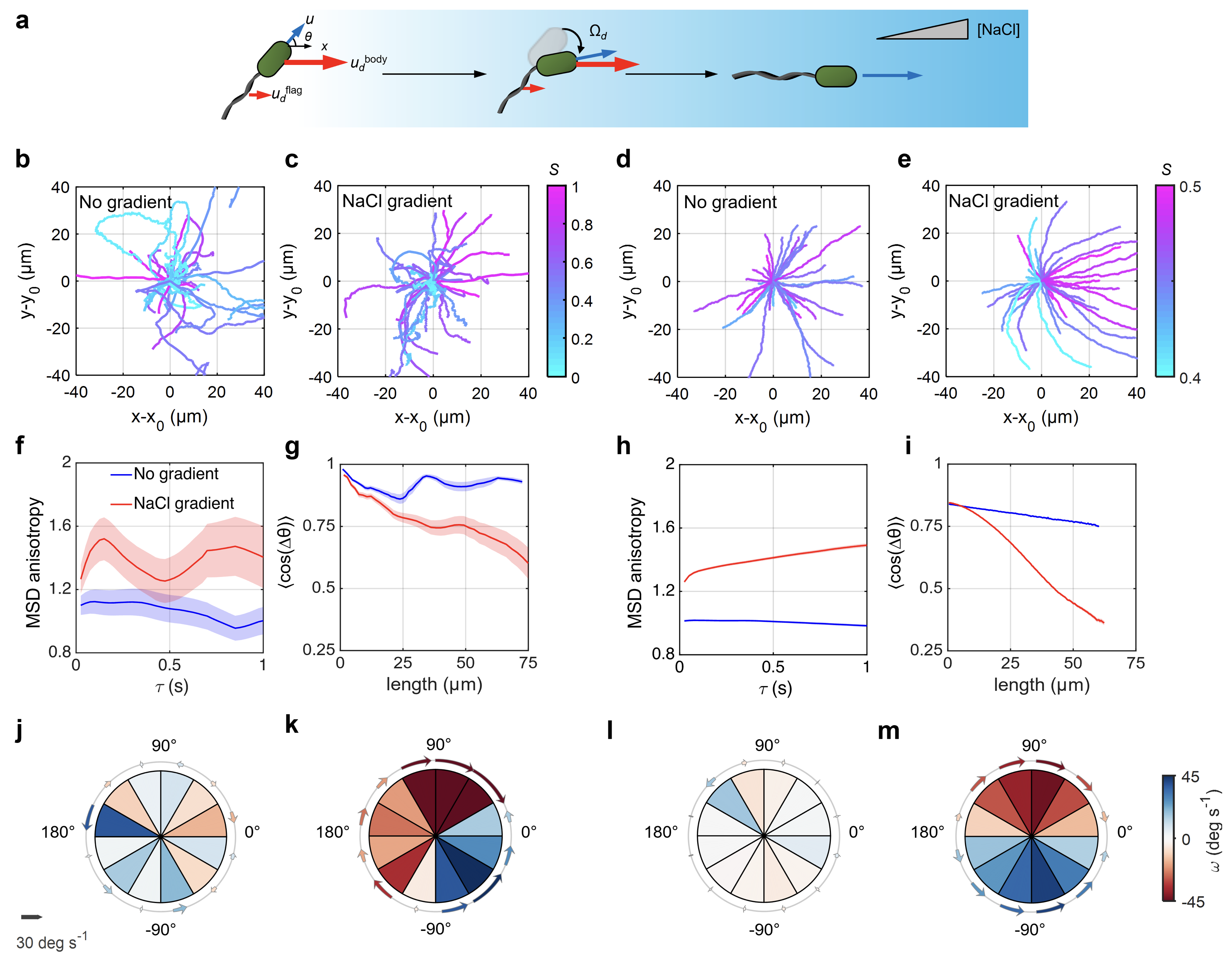}
	\caption{
		\textbf{Non-uniform diffusiophoresis steers \textit{P. putida}  toward salt.}	
		(a) An illustration of cell rotation due to non-uniform diffusiophoresis.
		(b,c) Experimental and (d,e) simulated run trajectories (40 randomly chosen) repositioned to start at the origin in the (b,d) absence (Movie S3,S5) and (c,e) presence (Movie S4,S6) of NaCl gradients. The color code represents run straightness $\mathcal{S}$.
		(f-i) Angle correlation function $\langle \cos(\Delta\theta) \rangle$ and MSD anisotropy ratio for (f,g) experimental and (h,i) simulated run trajectories.
		Shaded regions indicate $\pm$  standard error of the mean (SEM). (j-m) Instantaneous angular velocity distribution extracted from (j,k) experiments and (l,m) simulations in the (j,l) absence and (k,m) presence of NaCl gradients. 
		Positive and negative values indicate counterclockwise and clockwise rotation, respectively, with arrows
		representing rotation direction and magnitude.
	}
\end{figure*}

Therefore, when a cell swims toward salt, such an asymmetric action of diffusiophoresis tilts the cell body to align with the gradient (Figure 4a). 
This biased steering restores orientation along the gradient, stabilizes the trajectory, and enhances run speed by producing straighter paths.

To assess whether the rotation due to diffusiophoresis can overcome Brownian rotation, we evaluate the rotational P\'eclet number, 
\begin{equation}
	\mathrm{Pe}_r = \frac{2\Delta u_d}{(a+f)D_r} \sim \frac{2\pi\eta \Delta \mathcal{M}_d}{3kT\ell}\cdot\frac{(a+f)^2}{ \ln{\left(2(a+f)/b\right)}-1/2},
\end{equation}
where 
$\Delta u_d =u_d^{\textrm{body}}-u_d^{\textrm{flag}}$ and $\Delta \mathcal{M}_d =\mathcal{M}_d^{\textrm{body}}-\mathcal{M}_d^{\textrm{flag}}$ are the diffusiophoretic velocity and mobility difference between the body and the flagella bundle, respectively, and $D_r=\frac{3kT}{\pi\eta}\cdot\frac{\ln\left(2(a+f)/b\right)-1/2}{(a+f)^3}$ is the rotational diffusivity of the body plus flagella bundle \citep{berg2004coli}. We estimate $\mathrm{Pe}_r\sim10$, suggesting that such a subtle diffusiophoretic motion is yet powerful enough to overcome Brownian rotation, leading to improved motility toward salt. 
As shown in Figures~4b,c (Movies~S3,S4), the run trajectories repositioned to start at the origin $(x_0,y_0)=(0,0)$ indeed show the steering of cells toward the salt with straighter runs, confirming the impact of salt gradients on cell motion.

Cell steering due to diffusiophoresis can be further characterized by correlating the change of heading angle $\Delta \theta$ over its arc length $s$ along individual run trajectories, i.e., $\langle \cos(\Delta \theta) \rangle_s$, where $\langle \cdot \rangle_s$ is the ensemble average over all segments pairs with equal arc length separation across entire run trajectories (SM).
As shown in Figure 4g, in the absence of salt gradients, the angle correlation decays slowly along its displacement, indicating persistent swimming along its initial direction within a run. 
In contrast, under salt gradients, the angle correlation decays faster, indicating increased directional decorrelation over its run path. This long-range decay is not attributed to any change in the rotational Brownian noise; rather, it arises from continuous, deterministic reorientation of the swimming direction enabled by the diffusiophoretic torque. Such forced steering introduces long-range curvature into the run trajectories, leading to a faster decay of angle correlation as the motion becomes globally biased toward the salt gradient.

Further analyzing the mean square displacement (MSD) of the run trajectories reveals that the observed long-range steering arises mainly from the change of cell motion along the gradient direction (SM). The MSD anisotropy, which is defined as the ratio of MSD components along the gradient direction (MSD$_\parallel$) and transverse to the gradient direction (MSD$_\perp$), i.e., $\mathrm{MSD}_{\parallel}/\mathrm{MSD}_{\perp}$, remains approximately unity across all lag times $\tau$ in the absence of salt gradients (Figure 4f). However, under salt gradients, the MSD anisotropy is substantially increased, suggesting that the directional persistence is enhanced only along the gradient axis. Also, the MSD anisotropy is large even at very short lag times, indicating that the reorientation driven by diffusiophoretic torque occurs continuously throughout the entire run. 

These trends can be further corroborated by angular velocity distributions. Following \cite{stehnach2021viscophobic}, we calculated the instantaneous angular velocity of the cell body during individual runs as a function of the cell heading angle relative to the salt gradient direction. 
Without a gradient, the angular velocity remains small and nearly independent of heading angle (Figure 4j). In contrast, with NaCl gradients, the angular velocity becomes strongly heading-dependent (Figure 4k). Its magnitude was greatest for headings nearly perpendicular to the gradient, where a steering torque is expected to be maximum, and smaller for headings nearly parallel to it. The sign of the response was consistent with rotation toward the up-gradient direction. Comprehensive statistical analysis, including permutation tests for the MSD anisotropy, also confirms the robustness of our findings (Table~S3, SM).

The diffusiophoresis-driven steering can also be simulated by solving Langevin equations for cell translation and rotation (SM). The rotation of a Brownian cell subject to diffusiophoresis can be written as $\dot{\theta}(t)=\Omega_d+\eta_r(t)$, where $\Omega_d = -\frac{2\Delta u_d}{a+f}\sin{\theta}$ is the rotation due to non-uniform diffusiophoresis, and $\eta_r(t)$ is the random Brownian rotation. The simulation results are shown in Figures~4d,e where NaCl gradients effectively steer the cells toward salt and enable straighter runs, displaying a qualitative agreement with the experimental trajectories shown in Figure~4c. The non-uniform diffusiophoresis also yields increased MSD anisotropy (Figure~4h) and faster angle decorrelation (Figure~4i). Furthermore, the simulations (Figure 4m) also show angular dependency of cell motion on heading. Our simulations also predict the change in the global (Figure 1b) and local (Figure S4) cell population under the impact of non-uniform diffusiophoresis, which is consistent with the experimental results shown in Figures 1b and 2b.

We note that hydrodynamic effects other than diffusiophoresis may also arise due to salt gradients, which may contribute to steering. One possibility is rheotaxis driven by diffusioosmosis-induced recirculating flow \citep{shin2016size}. We estimate a diffusioosmotic wall shear rate of $\dot\gamma_\mathrm{wall} \sim 3\mathcal{M}_w/h\ell \sim 0.1$~s$^{-1}$ (SM), where $\mathcal{M}_w$ is the diffusioosmotic mobility of the wall and $h$ is the half-depth of the channel. This estimate is well below the shear rate associated with substantial rheotactic drift ($\sim 10$~s$^{-1}$) in the system studied by \cite{marcos2012bacterial}. 
Viscotaxis is likewise implausible, as the viscosity contrast between 1~mM and 100~mM NaCl is only about 1\% at room temperature, which is much smaller than the multi-fold contrast required for viscophobic turning examined by  \cite{stehnach2021viscophobic}. These comparisons suggest weak rheotactic and viscotactic contributions under our conditions. Taken together, the genetic control, heading-dependent turning, measured surface properties, and physical estimates support non-uniform diffusiophoresis as the mechanism most consistent with the observations among those considered.

\begin{figure}[t!]
\centering
\includegraphics[width=1.00\linewidth]{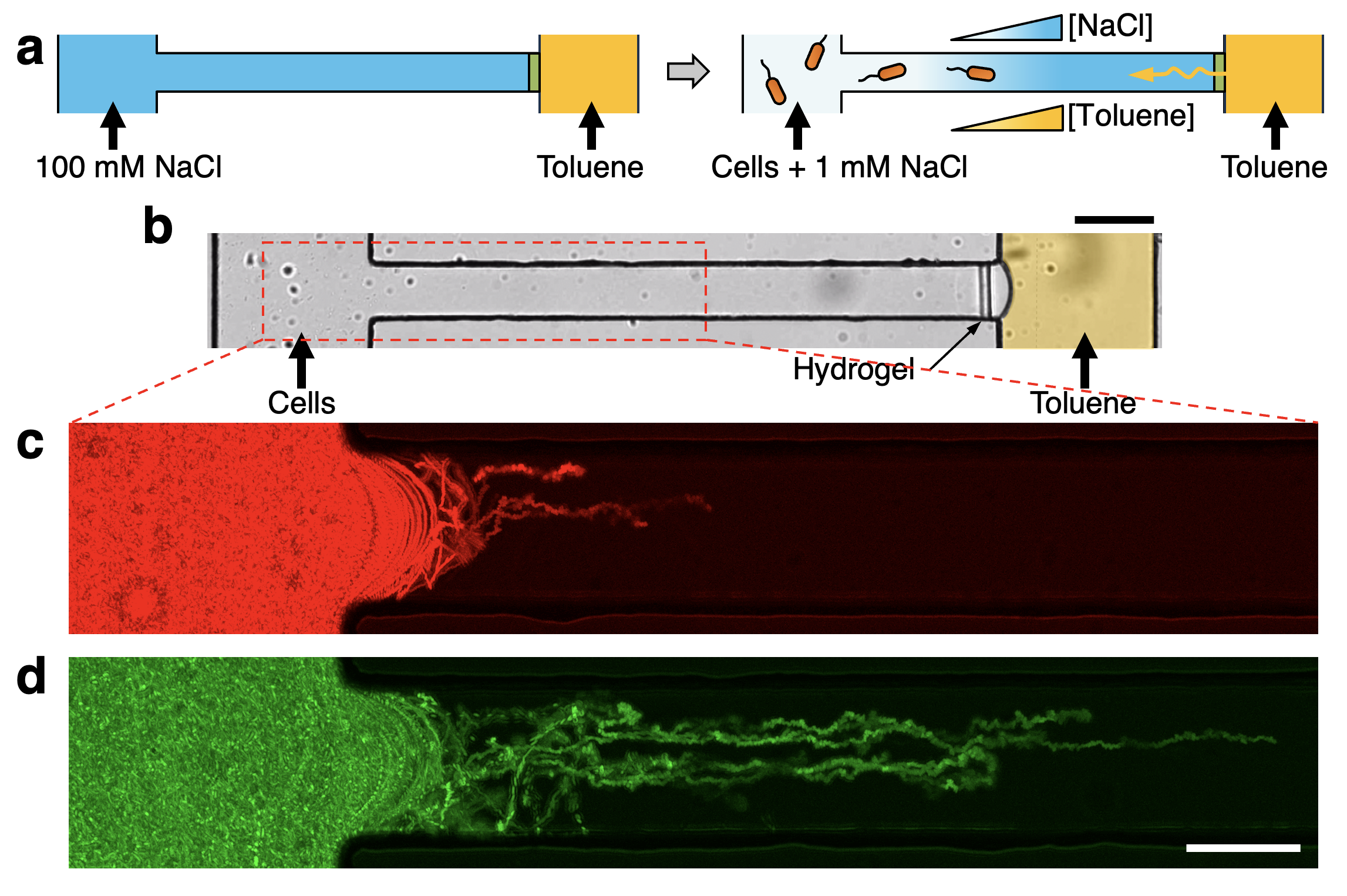}
\caption{
	\textbf{Salt gradients disperse cells toward toxic contaminant.}	
	(a) Simulating diffusiophoretic bioaugmentation of toluene-contaminated pore. Cells suspended in low salinity water are injected into the channel that is filled with high salinity water in the presence of toluene on the gel-sided channel, thereby creating dual chemical (salt and toluene) gradients. 
	(b) An image of the microfluidic channel in the presence of toluene. Toluene is false-colored. 
	(c,d) Trajectories of cells in the (c) absence (Movie S7)  and (d) presence of NaCl gradients. (Movie S8) 
	Scale bars in (b,d) are 50~$\mu$m. 
}
\end{figure}

\subsection*{Diffusiophoresis enhances bacterial dispersion toward pollutants}
Finally, we examined the diffusiophoretic cell dispersion in the presence of NAPL; toluene. 
While toluene is highly toxic to most microorganisms due to its membrane-disrupting effects, \textit{P. putida} is resistant and capable of metabolizing toluene, which drives its natural chemotaxis toward the compound \citep{zylstra1988toluene,inoue_pseudomonas_1989,ramos2002mechanisms,parales2000toluene}.
Using the same microfluidic device, we inject \textit{P. putida}  suspended in low salinity water (1~mM NaCl) into a channel that is initially filled with high salinity water (0.1~M NaCl) while toluene is present in the gel-sided flow channel (Figures~5a,b), effectively simulating diffusiophoresis-enabled bioaugmentation of NAPL-contaminated subsurface. Toluene gradually dissolves and creates toluene gradients in addition to salt gradients, creating dual chemical gradients. 

In the presence of toluene, we observe that cells initially chemotax toward toluene, but soon after they slow down due to the toxicity at high toluene concentration \citep{singh2010kinetics,wang2016enhanced}, making it challenging to disperse cells closer to the contaminant source (Figure~5c, Movie~S7). On the other hand, when NaCl gradients are additionally imposed, the cells continuously migrate into the pore via diffusiophoresis despite the lack of flagellar motility, as diffusiophoresis is forcing the cells to move toward the contaminant source (Figure~5d, Movie~S8). This demonstration suggests the potential utility of diffusiophoresis for bioremediation of NAPL-contaminated soil using externally imposed salt gradients.

\section*{Discussion}
Our results show that salt gradients can bias the directional motility of \textit{P. putida} in a confined pore. 
The $\Delta\textit{cheA}$ control demonstrates that the salt gradient response can occur without canonical CheA-dependent chemotaxis. The angular velocity also exhibits dependence on swimming heading consistent with a restoring torque.
Together with measured body/flagellar surface properties and Langevin simulations, these observations support differential diffusiophoresis of the cell body and flagellar bundle as the mechanism most consistent with the data. This process offers a potential strategy for bioremediation in contaminated soil environments, as demonstrated in Figure~5.

While this work only focuses on \textit{P. putida}, we expect other similar lophotrichous cells to experience similar biased motility due to the asymmetric geometry of the cell and the native surface charge of bacterial membranes  \citep{doan2020trace}.
In the same reasoning, we expect that peritrichous cells (e.g., \textit{Escherichia coli}) are likely to experience the diffusiophoresis-induced steering much less since their body and flagella are not spatially separated, unlike lophotrichous cells. 
This warrants further investigations to uncover how flagellar configuration determines cellular diffusiophoresis.

Our findings also connect to other physical mechanisms of microbial reorientation. In bacterial rheotaxis of \textit{Bacillus subtilis}, shear acting on the cell and helical flagellar bundle generates a torque that redirects the cell independent of chemical sensing \citep{marcos2012bacterial}. Viscosity gradients similarly drive viscophobic turning in the biflagellated alga \textit{Chlamydomonas reinhardtii}, for which uneven viscous loading of the two flagella produces rotation down the viscosity gradient \citep{stehnach2021viscophobic}. That response was characterized by a sinusoidal dependence of angular velocity on heading, which resembles what we report here for diffusiophoresis (e.g., Figure 4j-m). The present work identifies salt gradients, independent of flow, viscosity, or chemotaxis, as a distinct physical route by which swimming cells can be steered and guided, broadening the set of known non-chemoreceptor mechanisms for directed microbial transport.

\section*{Methods and Materials}

\subsection*{Materials}
Potassium phosphate dibasic (K$_2$HPO$_4$), potassium phosphate monobasic (KH$_2$PO$_4$), sodium chloride (NaCl), ethylenediaminetetraacetic acid (EDTA), toluene (99.8\%), and polyethylene glycol diacrylate (PEGDA; molecular weight = 700~Da) were purchased from Sigma-Aldrich. Polydimethylsiloxane (PDMS) was purchased from Dow Corning. Norland Optical Adhesive epoxy (NOA-81) was purchased from Norland Products Inc. 

\subsection*{Bacterial strains and growth conditions}
{\it{Pseudomonas putida}} F1 was kindly provided by Dr. Roseanne Ford at the University of Virginia ~\citep{ford2007role}. We used an aliquot of 50 \textmu l of frozen stocks of bacteria to inoculate 50 mL of Luria-Bertani (LB) broth. To induce chemotaxis, 15 \textmu l of 99.8\%\ toluene was added to the growth media. Cultures were grown aerobically on an orbital shaker (Ohaus) at 100~rpm and 28\textdegree C. Subsequently, cells were harvested when the optical density (\(\mathrm{OD}_{600}\)) reached 0.6, indicating mid-exponential growth phase. Harvested cells were then centrifuged and resuspended in 10$\times$ diluted random motility buffer (11.2~g/L K$_2$HPO$_4$, 4.8~g/L KH$_2$PO$_4$, and 0.029~g/L EDTA) for the motility study.  

\subsection*{Flagella separation and zeta potential measurements}
We employed the technique developed by Ali et al. \citep{ali2017bacteria} to isolate bacterial flagella for zeta potential measurements. {\it{P. putida}} cells were harvested at mid-log phase from the culture media, washed, and resuspended in phosphate buffer. Then flagella were detached from the cell bodies by vigorously vortexing the suspension for 10 min, followed by centrifugation at 10,000×$g$ to pellet the cell bodies. The supernatant containing flagella was centrifuged at 100,000×$g$ at 4°C to separate flagella from the buffer. The isolated flagella were subsequently resuspended in a 10 mM NaCl solution for zeta potential measurements. The zeta potential of cell bodies and flagella was determined using a dynamic light scattering instrument (Litesizer 500, Anton Paar).

\subsection*{Device design and fabrication}
The microfluidic channels were fabricated using a UV-curable epoxy, NOA-81 \citep{bartolo2008microfluidic}. Initially, a PDMS mold was made from a SU-8 master using standard soft lithography techniques. Subsequently, a few droplets of NOA-81 epoxy were added onto the PDMS mold and covered with a microscope glass slide. After partial curing of the epoxy under UV light, a second flat piece of partially cured NOA-81 on a thin glass slide was aligned and placed on top of the first glass slide. Then, the assembly was fully cured under UV to permanently bond the layers, forming a microfluidic channel with a channel height of 40~µm. 
The width and the length of the connecting pore were 65 by 800 µm.

The channels were then kept at 75\textdegree C for three days following UV curing to minimize residual toxicity of the epoxy to bacteria. Subsequently, a 20\%\ solution of PEGDA with a molecular weight of 700 is injected into the channels to form hydrogel membranes within the region of interest using localized UV light under an inverted microscope (DMi8, Leica). Then, the channels were washed for over 24 hours with DI water to remove any polymer residues from the channels. 

\subsection*{Microscopy and cell tracking}
The migration of bacterial cells was recorded at 40 fps under an inverted microscope equipped with an sCMOS camera (ORCA-Flash4.0 LT3, Hamamatsu). The position of bacteria was tracked using ImageJ with Trackmate plug-in. The data was then analyzed using MATLAB. For the run analysis, trajectories that span at least 10 frames (0.25 s) with a minimum displacement of 10 µm were considered. Each filtered trajectory was then smoothed using a 10-frame moving average, and both instantaneous and average velocities were calculated. Trajectories with velocities below 0.5 µm/s or above 200 µm/s were also removed. Moreover, trajectories exhibiting strong circular or highly curved motion, indicative of surface interactions near the top or bottom of the channel \citep{lauga2006swimming}, were also excluded from run analysis. Finally, any remaining outliers that were not automatically detected were manually removed from our run speed analysis.  To determine tumble events, we adopted the criteria developed by Alirezaeizanjani et al \citep{alirezaeizanjani2020chemotaxis}. In short, tumbles appear when there is a rapid change in the turning angle $\Delta\theta$, such that $\Delta\theta > \beta \sqrt{2D_r(t_2-t_1)}$, where $\beta=5/\sqrt{2}$ and $D_r = 0.1\ \text{s}^{-1}$. The tumbles can also occur when there is a sudden drop in the velocity $\Delta v/v_{min} >2$, where $v_{min}$ is the minimum velocity in the scanning period and $\Delta v$ is the maximum drop in the velocity between neighbor time $v(t_1)$ and $v(t_2)$, defined as $\Delta v = \mathrm{max}\left\{v(t_a) - v_\mathrm{min}, v(t_b)- v_\mathrm{min}\right\}$. A sample trajectory is shown in Figure S1.

\section*{DATA AVAILABILITY}
All codes, data files, and documentation for tracking can be accessed via the GitHub repository \url{https://github.com/alinikkh/-Pseudomonas-putida-F1-tracking}

\section*{Acknowledgments}
This material is based upon work supported by the National Science Foundation under Grants No.~2223737 and 2237177. 
We thank Drs. Roseanne Ford and Rhea Braun for providing \textit{P.~putida} strains.

%


\end{document}